\documentclass[12pt]{iopart}
\usepackage{graphicx}
\usepackage{color}
\usepackage{color}
\usepackage{epsfig}
\usepackage{graphicx}
\usepackage{bbm}
\usepackage{amsbsy}
\usepackage{slashed}
\usepackage{iopams}

\usepackage{xcolor}

\newcommand{\beq}{\begin{equation}}
\newcommand{\eeq}{\end{equation}}

\definecolor{mbcol}{rgb}{1,0,1}

\begin{document}
\title[IMC and Confinement within a Contact Interaction Model for Quarks]{Inverse magnetic catalysis and confinement within a contact interaction model for quarks}
\author{A Ahmad$^{1,2}$, A Raya$^1$} 
\address{$^1$Instituto de F\'{\i}sica y Matem\'aticas,
Universidad Michoacana de
San Nicol\'as de Hidalgo. Edificio C-3, Ciudad Universitaria, Morelia 58040, Michoac\'an, M\'exico.}
\address{$^2$Department of Physics, Gomal University, 29220 D.I. Khan, K.P.K., Pakistan.}

\ead{aftabahmad@ifm.umich.mx,raya@ifm.umich.mx}

\begin{abstract}
We evaluate the impact of an external magnetic field on the chiral symmetry and confinement-deconfinement transition temperatures by using a  vector-vector contact interaction model for  quarks regularized so as to include an explicit confining scale in the corresponding gap equation. Exploring the evolution of the chiral condensate and the confining scale with temperature $T$ and magnetic field strength $eB$ ($e$ represents the fundamental electric charge), we determine the pseudo-critical temperatures for the chiral ($T_c^\chi$) and deconfinement ($T_c^c$) transitions from their inflection points, respectively. By construction, $T_c^\chi= T_c^c$ in the chiral limit. Within a mean field approximation, we observe the magnetic catalysis phenomenon, characterized by a rising behavior of $T_c^\chi$ and $T_c^c$  with growing $eB$. Considering a lattice inspired running coupling which monotonically decreases with $eB$, inverse magnetic catalysis takes place in our model. We explore the role of the magnetic field in the traits of the confinement-deconfinement transition described by the model. Our findings are also in agreement with predictions derived from effective models of strong interactions.
\vspace{1pc}

\end{abstract}

\noindent{Published in: }{\jpg}{43 (2016) 065002}

\maketitle

\section{Introduction} \label{section-1}
Understanding the  different phases of  hadronic matter under extreme conditions is a major topic with implication in several branches of physics. Transition from a hadron gas to a quark--gluon plasma observed in heavy ion collisions can be understood from chiral symmetry breaking--restoration and/or confinement--deconfinement phase transitions.  Chiral symmetry breaking and confinement are two features of  low energy quantum chromodynamics (QCD), and thus cannot be described in the asymptotically free domain. Non-perturbative tools have been employed to explore these phenomena. Lattice simulation~\cite{Lattice}, Schwinger-Dyson Equations (SDEs)~\cite{SDE1, SDE2, SDE3, SDE4,sumrule} and other effective models of strong interactions~\cite{pNJL, Chqm, nNJL, NJL, Ayala:2015}  provide modern means to sketch the QCD phase diagram, which is usually drawn from the critical behavior of the respective (approximate) order parameters. 
For massless quarks, the chiral quark condensate $-\langle \bar{q}q\rangle^{1/3}$ is an order parameter for the chiral transition; it is finite at the chirally broken phase and vanishes at the critical temperature $T_c^\chi$. When finite current quark masses are considered, one can still determine $T_c^\chi$ --which is regarded as the pseudo-critical transition temperature for chiral restoration--from the inflection points of the thermal gradient of the condensate. Identifying an appropriate order parameter for confinement is less clear, however. The Polyakov loop~\cite{PolykovLoop:1978} and its dressed variant~\cite{Bilgici:2008} have been proposed to this end; their vanishing value at the confining phase rises maximally near the (pseudo-critical) transition temperature $T_c^c$, where deconfinement takes place. Nambu-Jona-Lasinio (NJL) model~\cite{Nambu:1961} has been extended to incorporate coupling of quarks to a homogeneous background gauge field representing the Polyakov loop and explore its impact in the chiral transition~\cite{Ratti}. Nevertheless, the static character of the Polyakov and dressed Polyakov loop might not be appropriate to describe the infra-red dynamics of light quarks dressed by a gluon cloud. Besides, these Polyakov loops lose their connection to confinement in field theories of strong interactions where the center symmetry of QCD is absent~\cite{Marquez:2015bca}. A more formal statement for this phenomenon  comes from the axiom of reflection positivity. The existence of inflection points in $n$-point Green functions is considered as the smoking gun of confinement~\cite{Roberts1}, although their physical interpretation still awaits.  Moreover, this argument is based on a nontrivial momentum dependence of the quark mass function, which in local models of the NJL type is not achievable. Nevertheless, one can still consider a propagator which does not develop poles hence describing an excitation that never reaches its mass shell. Physically, this corresponds to avoid quark production thresholds~\cite{Hugo:96}, rendering the NJL models one step closer to the actual description of quarks in the full QCD framework.

The QCD phase diagram becomes richer when we consider additional thermodynamical variables, like external fields. It is well known that strong magnetic fields have a tremendous impact in various physical systems.   
A typical example in astrophysics is a magnetar, in which the magnetic field might at the surface reaches intensities of the order of $B\sim 10^{10}$ Tesla~\cite{Duncan:1992}. As an estimated guess, a magnetic field  with intensity of the order of ~$B\sim 10^{19}$ Tesla was present during the electroweak phase transition and  $B\sim 10^{14}$ Tesla during the QCD phase transition~\cite{Vachaspati:1991nm, Enqvist:1993np}. On more terrestrial grounds, in non-central heavy ion collisions at RHIC and LHC, the generated magnetic fields are approximately of the order of~$B\sim 10^{14}$~-~$10^{16}$ Tesla~\cite{Skokov:2009qp} in intensity. A number of interesting effects are triggered by strong magnetic fields in QCD.
Among others, the chiral magnetic effect~\cite{Kharzeev} has attracted attention to explore topological features of vacuum and the  strong CP problem and has been recently measured in ZrTe$_5$~\cite{ZrTe5}. Moreover, magnetic fields are of direct relevance to understand the chiral and confinement phase transitions. It is known that a uniform magnetic field induces a dimensional reduction for charged Dirac fermions. Thus, fermion and antifermion pairs are closer together on the average, facilitating the formation of a chiral condensate, the so-called \textit{magnetic catalysis} effect~\cite{ 
Klimenko,magcat} (see~\cite{Shovkovy} for a recent review). This phenomenon is universal and  has been explored in NJL models~\cite{klev,klimenko,ferrer,Inagaki,Semenoff,Shi,chinese}, QED~\cite{qed1,qed2,qed3,qed4,qed5,qed6,qed7} and QCD~\cite{kar,miransky,mueller} among other theories and under a variety of external circumstances like a thermal bath, in low and high dimensions, finite density and so on. In QCD, it is observed an increase of the pseudo-critical transition temperatures as the strength of the magnetic field grows bigger~\cite{Buividovich:2010}. Nevertheless, because the dynamical mass generated by the magnetic field is much smaller than the constituent quark mass in vacuum for a wide range of magnetic field intensities~\cite{Shovkovy}, a field of this kind also produces a screening effect on gluon interactions in the infra-red, as can be accounted for from lattice~\cite{Bali:2012zg} and suggested by effective model calculations~\cite{Farias:2014, Ayala:2014gwa, Ferreira:2015gxa}. This phenomenon has been dubbed as \textit{inverse magnetic catalysis} and is responsible for a decreasing behavior of $T_c^{\chi,c}$ as the strength of the magnetic field increases. It is understood by realizing that being closer together, quark and antiquark pairs are reaching the asymptotic freedom regime faster by reducing the interaction strength as the intensity of the magnetic field increases~\cite{ayalaetal}. 
Inverse magnetic catalysis is observed to take place for magnetic fields of arbitrary intensity so long as $T\ge T_c^{\chi,c}$ and persists for very strong magnetic fields,  changing the behavior of the confinement-deconfinement transition from a cross over to a first order one~\cite{Edrodi} (see Ref.~\cite{reviewPDMF} for a recent review of the magnetized QCD phase diagram).
 In this work, we use a confining variant of the NJL model regularized within a proper-time scheme~\cite{Hugo:96} to study the effect of external magnetic field on the chiral condensate and the confining scale. This vector-vector \textit{contact interaction} model has been successfully used to reproduce hadronic static properties of pions and other low energy mesons and baryons in vacuum~\cite{Xiomara-2010}. Extensions of this model at finite temperature~\cite{Wang:2013} or density~\cite{Klahn} have already been considered. We take a step forward and include a uniform magnetic field in the formalism. In the mean field approximation, our model exhibits magnetic catalysis only, whereas improving the approximation by including a magnetic field dependence on the coupling, we find signals of inverse magnetic catalysis, providing further support to the ideas of 
 Ref.~\cite{Farias:2014, Ayala:2014gwa, Ferreira:2015gxa,ayalaetal}.  
The paper is organized as follows: In Sec.~2, we discuss the gap equation for the contact interaction  model at zero and finite temperature. Entanglement between dynamical chiral symmetry breaking and confinement is expressed through an explicit temperature dependent regulator in the infra-red. In Sec.~3, we add the influence of an external magnetic field by using the Schwinger proper-time representation of the quark propagator. Zero and finite temperature are considered and the numerical solution of the gap equation is discussed. Summary and conclusion are presented in Sec.~4.

\section{Contact Interaction} \label{section-2}

We start our discussion by presenting the generalities of the Contact Interaction Model of QCD to put in a broader perspective the problem we address in this work. In QCD, quarks interact via vector-boson exchange and the 
SDE for the quark propagator is
\begin{eqnarray}
S^{-1}(p)&=&S_0^{-1}(p)
+\int \frac{d^4k}{(2\pi)^4} g^2
\frac{\lambda^a}{2}\gamma_\mu S(k)
\frac{\lambda^a}{2}\Gamma_\nu(k,p)\Delta_{\mu\nu}(k-p)\,,
\label{gap-QCD}
\end{eqnarray}
where $S(p)$ is the full quark propagator and $S_0(p)$ its bare
counterpart, $g$ is the coupling constant, $\lambda^a$ are the
Gell-Mann matrices and $\Gamma_\nu(k,p)$ and
$\Delta_{\mu\nu}(k-p)$ represent, respectively, the full
quark-gluon vertex and the gluon propagator.  In the low momentum regime, relevant to explore non perturbative features of strong interactions, gluons acquire a dynamical mass, as has been accounted for in lattice studies~\cite{GluonMass}. This opens the possibility of describing QCD in an effective manner through NJL type of interactions from the Lagrangian
\begin{equation}
{\cal L}=\bar\psi (i\not \! {\partial}-m_0)\psi + \frac{G}{2}[(\bar\psi\psi)^2+(\bar\psi i\gamma_5 \vec{\tau}\psi)^2],\label{NJL}
\end{equation}
where the four-Fermi interactions term contains a scalar and an axial-vector interaction piece ($\tau$ representing the Pauli matrices in isospin space) and $G$ is the coupling of the theory.
Such a Lagrangian describes spontaneous chiral symmetry breaking from the gap equation
\begin{equation}
\frac{M-m_0}{2G}=-{\rm Tr}\int^\Lambda \frac{d^4k}{(2\pi)^4} S(k)\;,\label{gapNJL}
\end{equation}
where $M$ is the dynamical mass and the symbol $\int^\Lambda$ stresses the need to regularize the integrals. Within the SDE formalism, it
  has been shown in
a series of articles at zero temperature that the static
properties of low energy mesons and baryons can be faithfully
reproduced by assuming that quarks interact not via massless
vector-boson exchange but instead through the following contact
interaction~\cite{Xiomara-2010}:
 \begin{eqnarray}
  g^2 \Delta_{\mu \nu}(q) = \delta_{\mu \nu} \frac{4 \pi
  \alpha_{\rm IR}}{m_G^2} \equiv \delta_{\mu \nu} \alpha_{\rm eff}(0),
  \label{CImodel}
 \end{eqnarray}
where $m_G=800$ MeV is a gluon mass scale which, as we earlier mentioned, is in fact
generated dynamically in QCD~\cite{GluonMass}, and $\alpha_{\rm
IR}=0.93\pi $ specifies the interaction strength in the infra-red (in Eq.~(4) of the first article in Ref.~\cite{Xiomara-2010} we identify $1/m_G^2 \to 4 \pi
  \alpha_{\rm IR}/m_G^2$ of our present conventions).  Written in this form, it is obvious that if the coupling $\alpha_{\rm IR}$ is small and and the gluon mass $m_G$ is large, there is a critical value of $\alpha_{\rm eff}(0)$ above which chiral symmetry is broken, but below this critical value, it is impossible in the model to generate masses dynamically.

 We  proceed to embed this interaction in a rainbow-ladder
truncation of the gap equation, Eq.~(\ref{gap-QCD}). 
Wavefunction renormalization becomes trivial in this model, and the 
quark mass function becomes momentum independent, namely, a constant which upon writing $d^4k=(1/2)k^2dk^2\sin^2\theta d\theta \sin \phi d\phi d\psi$, after performing the trivial angular integrations and changing variables as $s=k^2$, we determine self-consistently from
\begin{eqnarray}
M=m_{0}+\frac{ \alpha_{\rm eff}(0) M }{3\pi^2}\int^{\infty }_{0}
ds\frac{s}{s+M^2} \,. \label{PTRa}
\end{eqnarray}
We emphasize that it has the same functional form as in Eq.~(\ref{gapNJL}). In order to regularize the integrals, we now exponentiate the
denominator of the the integrand and employ the confining
proper-time regularization~\cite{Hugo:96},
\begin{eqnarray}
\frac{1}{s+M^2}=\int^{\infty }_{0} d\tau {\rm e}^{-\tau(s+M^2)}
\rightarrow \nonumber\\
\int^{\tau_{ir}^2}_{\tau_{uv}^2} d\tau {\rm
e}^{-\tau(s+M^2)}
=\frac{ {\rm
e}^{-\tau^{2}_{uv}(s+M^2)}-{\rm e}^{-\tau^{2}_{ir}(s+M^2)}}{s+M^2}
\,. \label{PTRb}
\end{eqnarray}
Here, $\tau^{-1}_{ir,uv}=\Lambda_{ir,uv}$ are respectively, the
infra-red and ultra-violet regulators. This procedure ensures the
absence of real as well as complex poles in the quark
propagator. The infra-red cut-off corresponds
to the confinement scale whereas the ultra-violet cut-off plays a
dynamical role due to the non-renormalizability of the model. The
pole-less structure of the quark propagator corresponds to the
absence of the quark production thresholds and it is another 
analytic form consistent with quark confinement~\cite{Hugo:96}; an excitation
described by a pole-less propagator would never reach its
mass-shell. The gap equation, after integration over $s$, can now be written as
\begin{eqnarray}
M &=& m_0 
+\, \frac{M^3 \alpha_{\rm eff}(0)}{3\pi^2}
\left[\Gamma(-1,M^2\tau_{uv}^2)-\Gamma(-1,M^2\tau_{ir}^2)\right]
\, , \label{PTRd}
\end{eqnarray}
where  
$$\Gamma(\alpha,x) = \int_x^{\infty} t^{\alpha-1} {\rm
e}^{-t} dt$$ 
is the incomplete Gamma function. We use the parameters of Ref.~\cite{Xiomara-2010}, namely, we fix the coupling to 
\begin{eqnarray}
       \alpha_{\rm eff}(0) &=& 5.739\cdot 10^{-5}~\mathrm{MeV^{-2}},\label{p1}
\end{eqnarray} 
and use the infra-red and ultra-violet cut-offs as
\begin{eqnarray}
       \tau_{ir} \quad=\quad (240~\mathrm{MeV})^{-1}, \qquad\qquad
       \tau_{uv} \quad=\quad (905~\mathrm{MeV})^{-1}
       \label{p2}
\end{eqnarray}
\%begin{eqnarray}
which have been fitted to vacuum properties in the pion and rho-meson sector.
With these parameters, considering a current quark mass of $m_{0}=7$ MeV, the constituent quark mass and
the chiral condensate per flavor are calculated to be $M=367$~MeV and
$\langle\bar{u}u\rangle^{1/3} = \langle\bar{d}d\rangle^{1/3} = -243$~MeV, respectively.

At finite temperature, in the imaginary time formalism, we split the fermion four-momentum according to $k=(\omega_{l},\vec{k})$, where $\omega_l = (2l+1)\pi T$ are the fermionic Matsubara frequencies and we adopt the standard convention for momentum integrations, namely,

\begin{eqnarray}
\int\frac{d^4k}{(2\pi)^4} \rightarrow T \sum_{n} \int\frac{d^3k}{(2\pi)^3}. \label{mf}
\end{eqnarray}
Thus, the gap equation~(\ref{gapNJL}) at finite temperature becomes
\begin{eqnarray}
M= m_0+\frac{8 \alpha_{\rm eff}(0) MT}{3\pi^2} \sum^{\infty}_{l=-\infty} \int_0^\infty d\vec{k} \vec{k}^2 \frac{1}{\vec{k}^2 +\omega^{2}_{l} + M^2},\label{gapT1}
\end{eqnarray}
 This equation and some of its variants have been discussed in several works~\cite{Wang:2013}. For our purposes, in order to ensure coincidence of chiral symmetry and confinement transitions, we regularize the integrals by exponentiating the denominator for each $\omega_l$, i.e.,
\begin{eqnarray}
\frac{1}{\vec{k}^2 +\omega^{2}_{l}+ M^2}\longrightarrow  \int^{\tilde{\tau}^2_{ir}}_{\tau^2_{uv}}d\tau {\rm e}^{-\tau(\vec{k}^2 +\omega^{2}_{l}+ M^2)} \label{gapT2},
\end{eqnarray}
with 
\begin{eqnarray}
\tilde{\tau}_{ir}=\tau_{ir} \frac{M(0)}{M(T)} \label{gapT3},
\end{eqnarray}
 such that  in the chiral limit $m_0=0$, the confining scale vanishes  (or, equivalently,  $\tilde{\tau}_{ir}\to\infty$) at the chiral symmetry restoration temperature, allowing poles in the propagator to develop when current quark masses are finite. This is a simple way of ensuring the coincidence between confinement and chiral symmetry transitions. Summation over Matusbara frequencies and the remaining radial integration can be performed from the identities
 
 \begin{eqnarray}
\sum^{\infty}_{l=-\infty} e^{-\tau\omega_{l}^2}=\Theta_{2}(0,e^{-4 \pi^2 \tau T^2})\;,\qquad
\int^{\infty}_{0} d \vec{k} \vec{k}^2 e^{-\tau\vec{k}^2}
= \frac{\sqrt{\pi}}{4 \tau^{3/2}},\label{gapT4}
\end{eqnarray}   
where $\Theta_{2}(x,y)$ represents the second Jacobi theta function. Then, we finally reach at the following expression for the gap equation
\begin{eqnarray}
M &=&m_0 + \frac{2M \alpha_{\rm eff}(0)T}{3\pi^{3/2}} 
\int^{\tilde{\tau}^2_{ir}}_{\tau^2_{uv}}d\tau \frac{ {\rm e}^{-M^2\tau}\Theta_{2}(0, {\rm e}^{-4\pi^2 T^2\tau})}{\tau^{3/2}}\label{gapT5},
\end{eqnarray}
\begin{figure}[t!]
\begin{center}
\includegraphics[width=0.48\textwidth]{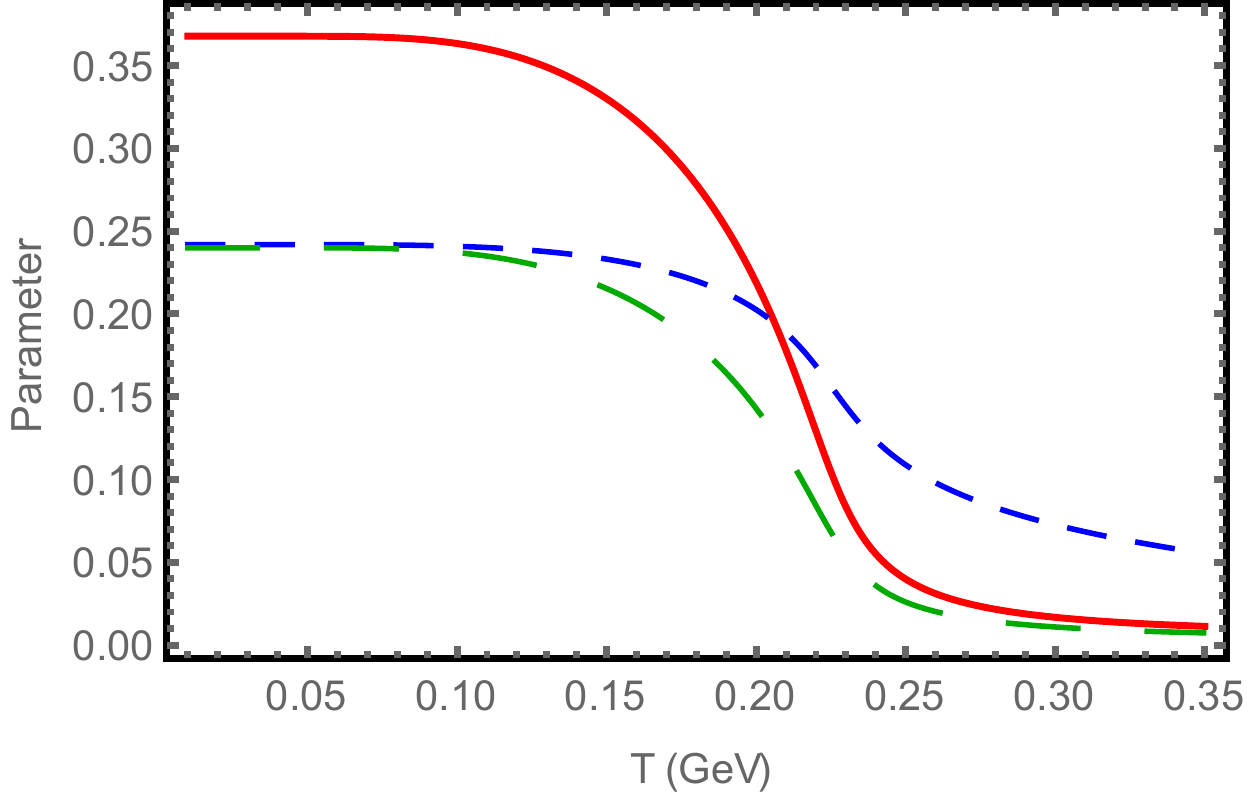}
\caption{ Dynamical mass (red, solid curve);  chiral condensate (blue, short-dashed curve) and  confining scale (green, long-dashed curve),  as functions of temperature for a current quark mass $m_{0}=7$~MeV.}
\label{Fig.1}
\end{center}
\end{figure}
\noindent
With a current quark mass $m_0=7$~MeV, the thermal evolution of the dynamical mass $M$, the chiral condensate $-\langle \bar{q}q\rangle^{1/3}$ and the confinement scale $\tilde{\tau}_{ir}^{-1}$ are shown in Fig.~\ref{Fig.1}. Their $T=0$ values correspond to those obtained from the parameters in Eqs.~(\ref{p1}) and~(\ref{p2}). 
\begin{figure}[th!]
\begin{center}
\includegraphics[width=0.48\textwidth]{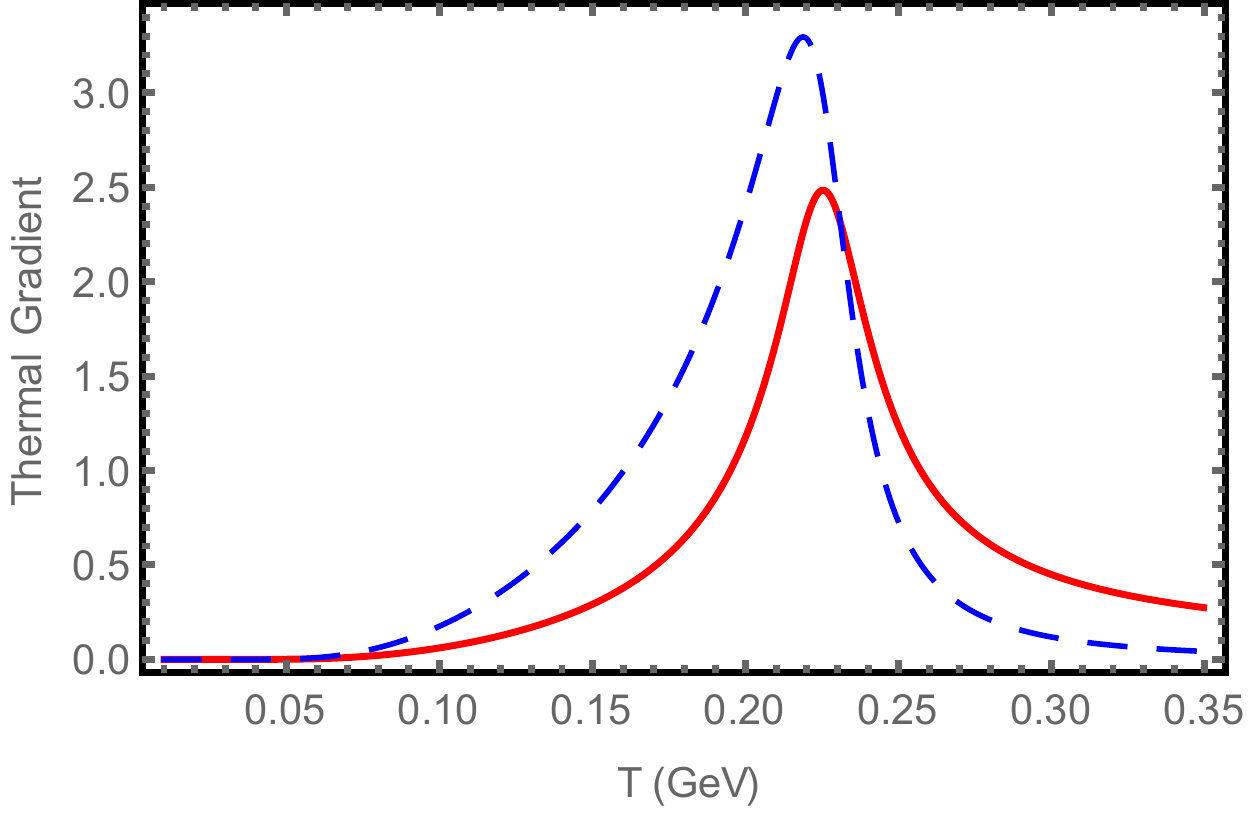}
\caption{Thermal gradient of the condensate $-\partial_T\langle \bar{q}q\rangle^{1/3}$ (red, solid curve) and of the confining scale (blue, dashed curve) $\partial_T \tilde{\tau}_{ir}^{-1}$,  as  functions of temperature for a current quark mass $m_{0}=7$~MeV. $T_c^\chi=T_c^c\equiv T_c \simeq 225$~MeV within numerical accuracy.}
\label{Fig.2}
\end{center}
\end{figure}
The pseudo-critical temperatures for the chiral symmetry breaking-restoration $T_c^\chi$ and confinement-deconfinement $T_c^c$ transitions are determined, respectively, from the position of the maxima of their thermal gradients $-\partial_T\langle \bar{q}q\rangle^{1/3}$ and $\partial_T\tilde{\tau}_{ir}^{-1}$, shown in Fig.~\ref{Fig.2}. 
We observe that within the numerical accuracy, both pseudo-critical temperatures are coincidental, $T_c^\chi=T_c^c\equiv T_c\simeq 225$~MeV. Next, we include the influence of a magnetic field within this framework.

\section{Gap equation in a Magnetic field}

Now, we consider a background homogeneous magnetic field directed along the $z$-axis, with magnitude $B$ and defined through the symmetric gauge vector potential

\begin{eqnarray}
A^{ext}_{\mu}=\left( 0,-\frac{B y}{2}, \frac{B x}{2},0\right) 
\end{eqnarray}

\begin{figure}[t!]
\begin{center}
\includegraphics[width=0.5\textwidth]{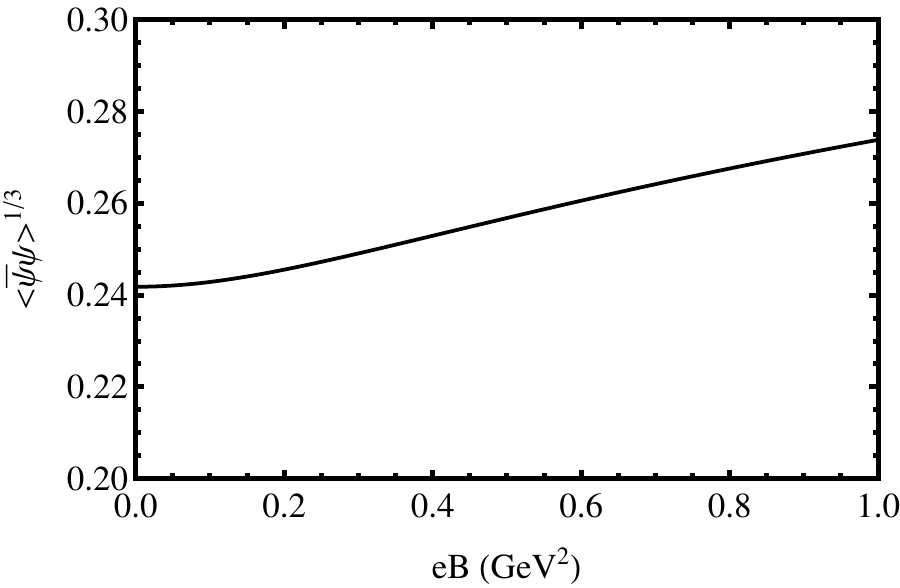}
\caption{Quark-antiquark condensate as a function of $eB$ at zero temperature. Plot shows that the quark-antiquark condensate increases with the increase of the magnetic field strength.}
\label{FMag1}
\end{center}
\end{figure}

 From the NJL point of view, the corresponding Lagrangian of the theory is generalized from the one in Eq.~(\ref{NJL}) as~\cite{vivian2}
\begin{equation}
{\cal L}=\bar\psi (i\not \! {D}-m_0)\psi + \frac{G}{2}[(\bar\psi\psi)^2+(\bar\psi i\gamma_5 \vec{\tau}\psi)^2]+\frac{G'}{2}[(\bar\psi\Sigma^3\psi)^2+(\bar\psi i\gamma_5 \vec{\tau}\Sigma^3\psi)^2],
\end{equation}
where $D_\mu=\partial_\mu -ieA_\mu^{\rm ext}$ and $\Sigma^3=(i/2)[\gamma^1,\gamma^2]$. The new term contains tensor interactions which emerge when Lorentz symmetry of the NJL Lagrangian is explicitly broken, and give rise to new condensates~\cite{vivian2} related to the coupling $G'$,  which is weak, $G'<G$, unless the magnetic field is very strong. In the very strong magnetic field regime, many interesting novel phenomena take place. For instance, anisotropies in the strong coupling constant~\cite{vivian1} develop. Nevertheless, we consider the situation where $G'$ is negligible. Thus, considering the scalar channel and expanding the operator $\bar\psi\psi$ around its vaccum expectation value, $\bar\psi\psi=\langle\bar\psi\psi\rangle+\delta \bar\psi\psi,$ such Lagrangian can be cast in the form
\begin{equation}
{\cal L}=\bar\psi (i\not \! {D}-M)\psi -\frac{(M-m_0)^2}{4G} +\ldots,
\end{equation}
that corresponds to the Lagrangian of a quark of (constant) mass $M$ in a magnetic field plus an irrelevant constant term. Hence, the gap equation continues to be of the form~(\ref{gapNJL}) 
where now $S(k)$ is dressed with magnetic field effects, $S(k)\to \tilde{S}(k)$ in the Fock-Schwinger representation~\cite{Schwinger:1951}, namely
\begin{eqnarray}
\tilde{S}(k)=\int^{\infty}_{0} ds \frac{{\rm e}^{-s(k^{2}_\parallel+k^{2}_{\perp}\frac{{\rm tan}(|q_{f}Bs|)}{|q_{f}Bs|}+M^{2})}}{{{\rm cosh}(|q_{f}B s|)}}\nonumber\\ \times \bigg[\bigg({\rm cosh}(|q_{f}Bs|)-i\gamma^{1}\gamma^{2}{\rm sinh}(|q_{f}Bs|)\bigg)
(M-\slashed{k}_{\parallel})-\frac{\slashed{k}_{\perp}}{{\rm cosh}(|q_{f}Bs|)} \bigg], \label{sh}
\end{eqnarray} 
where the parallel and transverse splitting of the quark momenta is in reference to the magnetic field direction, as usual~\footnote{Recall that $k^2=k_\parallel^2+k_\perp^2$, with $k_\parallel^2= k^{2}_0+k^{2}_{3}$ and $k_\perp^2= k^{2}_2+k^{2}_{4}$}, and  $q_{f}=(q_u=+2e/3 ,q_d=-e/3)$ refers to the electric charges of up and down quarks.  With these  ingredients, adopting the regularization procedure of the previous section,
the corresponding gap equation for the  dynamical mass at zero temperature under the influence of a uniform magnetic field for each light quark flavor becomes

\begin{eqnarray}
M_{u,d}&=& m_{0}+\frac{16 \alpha_{\rm eff}(0) M_{u,d}}{3} \int \frac{d^2k_{\perp}}{(2\pi)^2}\frac{d^2k_{\parallel}}{(2\pi)^2}\nonumber\\&&\times 
\int^{\tau^{2}_{ir}}_{\tau^{2}_{uv}} d\tau {\rm e}^{-\tau(k^{2}_{\parallel}+k^{2}_{\perp}\frac{{\rm tanh}(|q_{u,d}B\tau|)}{|q_{u,d}B\tau|}+M_{u,d}^{2})}.\label{Mag7}
\end{eqnarray}
Making  use of the relations
\begin{eqnarray}
 \int \frac{d^2 k_{\parallel}}{ (2\pi)^2}e^{-\tau k^{2}_{\parallel}} = \frac{1}{4\pi \tau},\qquad \int \frac{d^2 k_{\perp}}{ (2\pi)^2} e^{-\tau k^{2}_{\perp} 
\frac{{\rm tanh}(|q_{f}B\tau|)}{|q_{f}B\tau|}} =\frac{|q_{f}B|}{4\pi{{\rm tanh}(|q_{f}B\tau|)}},
\end{eqnarray}
the expression for the gap equation for  the average dynamical mass 
\begin{equation}
M=\frac{1}{2}(M_{u}+M_{d})
\end{equation}
for light quarks at zero temperature in a magnetic field is of the form
\begin{eqnarray}
M &=&m_{0}+ \frac{ \alpha_{\rm eff}(0)}{3\pi^{2}} \sum_{f=u,d} |q_{f}B| \int^{{\tau}^2_{ir}}_{\tau^2_{uv}}d\tau \frac{{M\rm e}^{-M^2\tau}}{\tau {\rm \tanh}(|q_{f}B|\tau)},\label{EMag2}
\end{eqnarray}
We solve the above gap equation with the constant $\alpha_{\rm eff}(0)$ of Eq.~(\ref{p1}). From the numerical results, we plot the average condensate,
\begin{equation}
-\langle\bar{\psi}{\psi}\rangle^{1/3}=-\frac{1}{2}(\langle\bar{u}{u}\rangle^{1/3}+\langle\bar{d}{d}\rangle^{1/3}),
\end{equation}
as a function of magnetic field strength in Fig.~\ref{FMag1}. We observe that the chiral quark condensate increases as we increase the magnetic field strength, the unequivocal sign of Magnetic Catalysis taking place. 

\begin{figure}[t!]
\begin{center}
\includegraphics[width=0.48\textwidth]{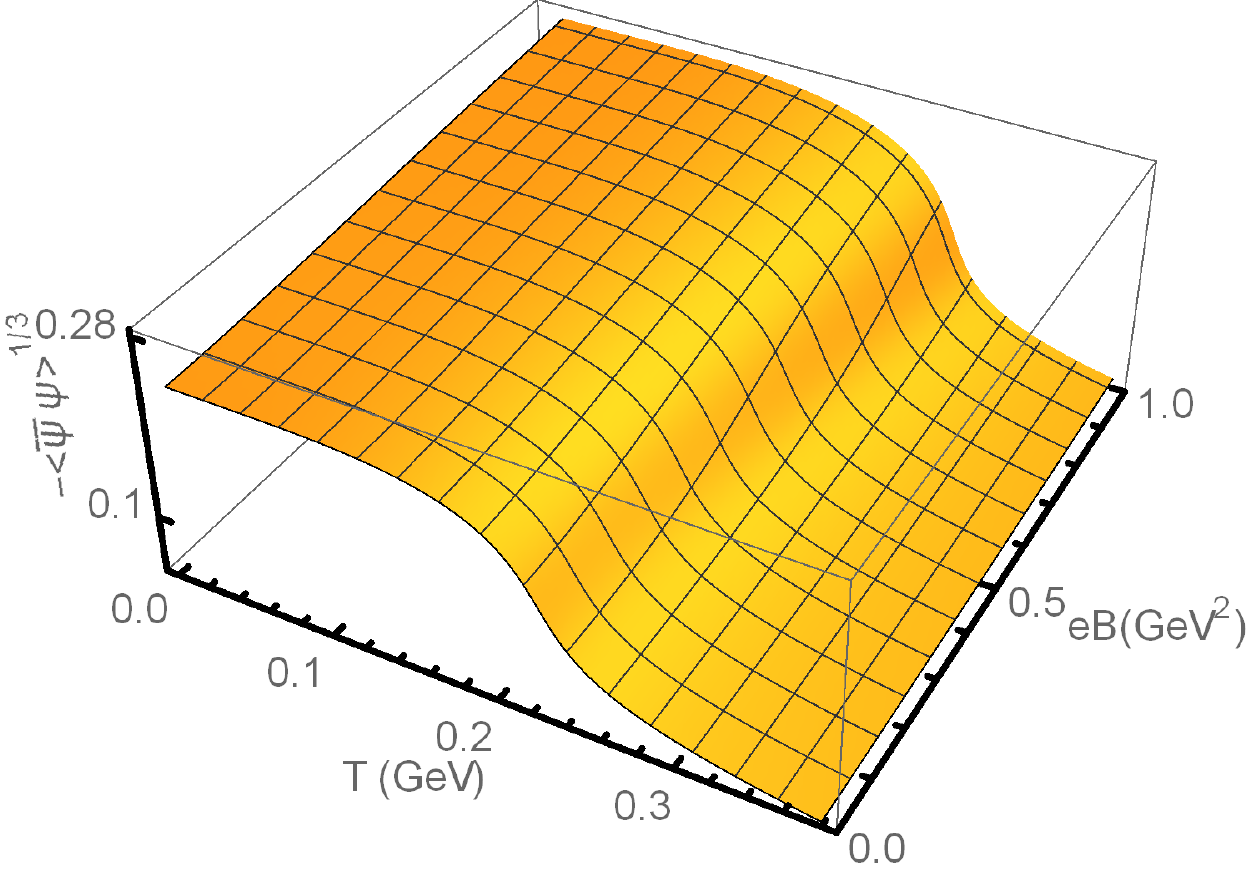}
\caption{ 
Average chiral condensate 
as a function of temperature and magnetic field strength with a current quark mass $m_{0}=7$~MeV. Plot is generated with the constant $\alpha_{\rm eff}(0)$ in Eq.~(\ref{p1}).}
\label{Fig.3}
\end{center}
\end{figure}
At finite temperature $T$ and under the influence of the magnetic field,  the gap equation now reads 
\begin{eqnarray}
M&=& m_{0}+ \frac{16  \alpha_{\rm eff}(0)}{3} M T\sum_{f=u,d}\nonumber\\
&&\times\sum^{\infty}_{l=-\infty}\int\frac{d k_{3}}{(2\pi)}\frac{d^{2} k_{\perp}}{(2\pi)^{2}} \int^{\tau^{2}_{ir}}_{\tau^{2}_{uv}} d\tau {\rm e}^{-\tau(\omega^{2}_{l}+k^{2}_{3}+k^{2}_{\perp}\frac{{\rm tan}(|q_{f}B\tau|)}{|q_{f}B\tau|}+M^{2})} \label{MagT3}.
\end{eqnarray}
As before, after performing the sum over Matsubara frequencies and  integrating over $k_3$ and $k_\perp$, the corresponding gap equation for  $M$ becomes  
\begin{eqnarray}
M &=&m_{0}+ \frac{2M \alpha_{\rm eff}(0)T}{3\pi^{3/2}} \sum_{f=u,d} |q_{f}B| 
\int^{{\bar{\tau}}^2_{ir}}_{\tau^2_{uv}}d\tau \frac{{\rm e}^{-M^2\tau}{\Theta_{2}(0,{\rm e}^{-4\pi^2 T^2\tau})}}{\tau^{1/2}\ {\rm \tanh}(|q_{f}B|\tau)},
\end{eqnarray}
where
\begin{eqnarray}
\bar{\tau}_{ir}=\tau_{ir} \frac{M(0,0)}{M(T,eB)}.
\end{eqnarray}
We solve the above gap equation with the constant $\alpha_{\rm eff}(0)$ of Eq.~(\ref{p1}). We plot the averaged condensate  as a function of temperature and magnetic field strength in Fig.~\ref{Fig.3}.
 We identify the temperature at which the thermal gradient of the chiral condensate peaks to specify $T_{c,B}^\chi$. Our scheme ensures that $T_{c,B}^\chi\simeq T_{c,B}^c\equiv T_{c,B}$.

\begin{figure}[t!]
\begin{center}
\includegraphics[width=0.48\textwidth]{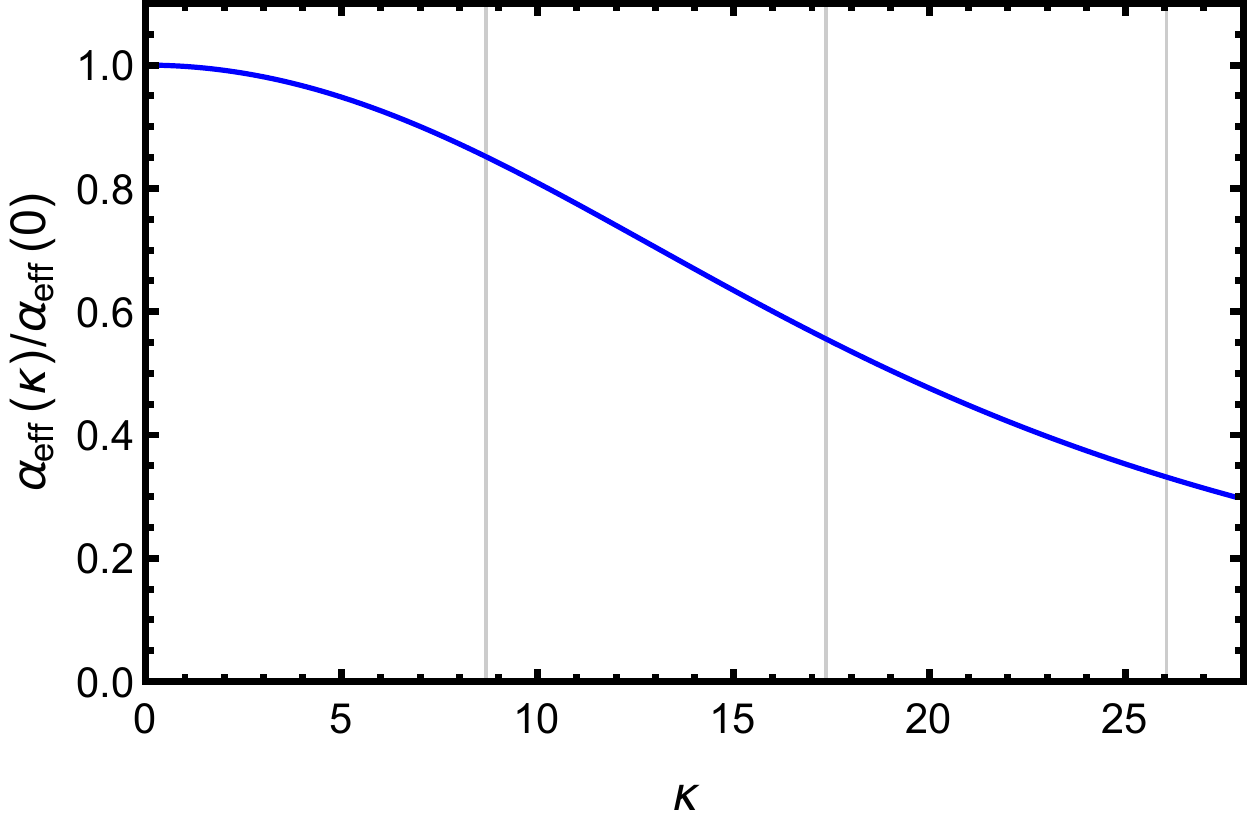}
\caption{Effective coupling $\alpha_{\rm eff}(\kappa)$ in Eq.~(\ref{coup}) normalized to the constant value of $\alpha_{\rm eff}(0)$ in Eq.~(\ref{p1}). Vertical lines, from left to right correspond to $eB=$0.5, 1 and 1.5 GeV$^2$.} 
\label{Fig.4}
\end{center}
\end{figure}

\begin{figure}[t!]
\begin{center}
\includegraphics[width=0.48\textwidth]{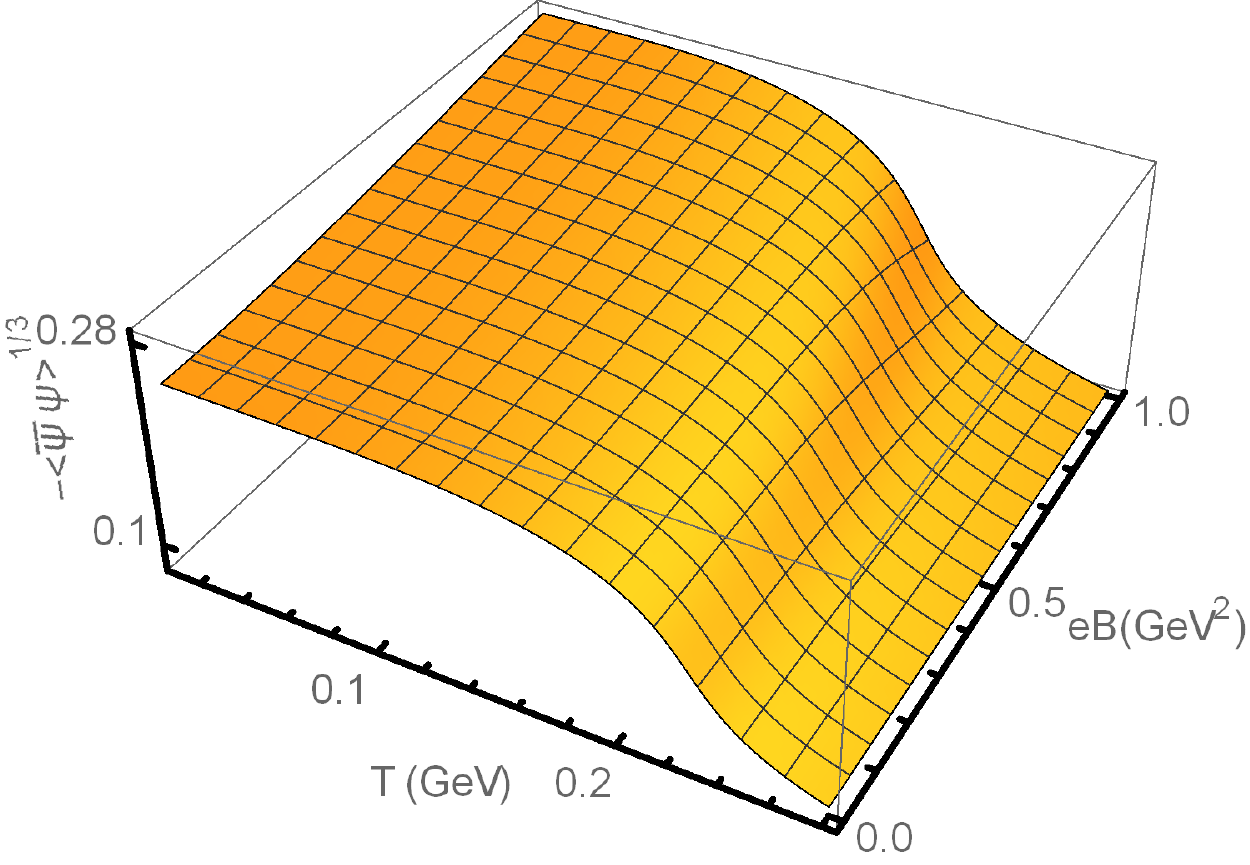}
\caption{
Average chiral condensate 
as a function of temperature and magnetic field strength for a current quark mass $m_{0}=7$~MeV. Plot is generated with  $\alpha_{eff}(\kappa)$ in Eq.~(\ref{coup}).} 
\label{Fig.5}
\end{center}
\end{figure}

\begin{figure}[th!]
\begin{center}
\includegraphics[width=0.48\textwidth]{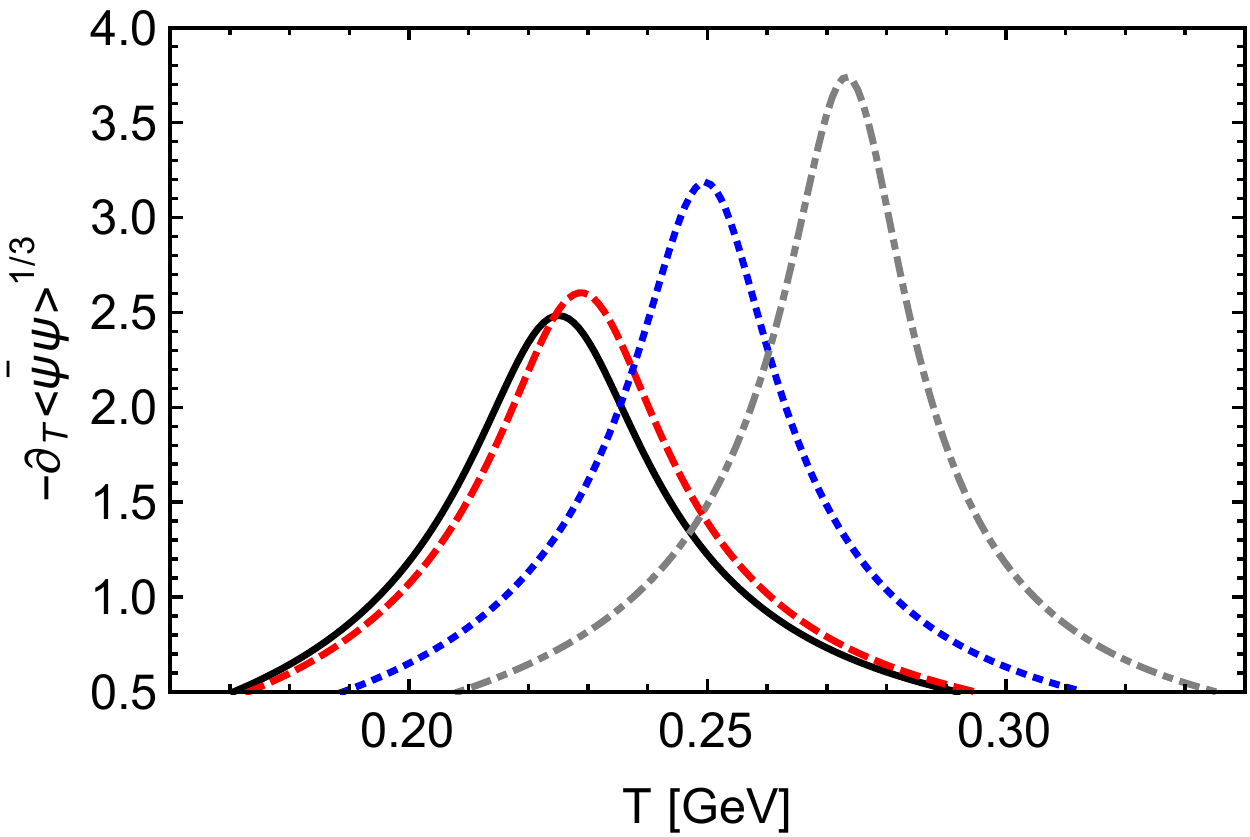}
\includegraphics[width=0.48\textwidth]{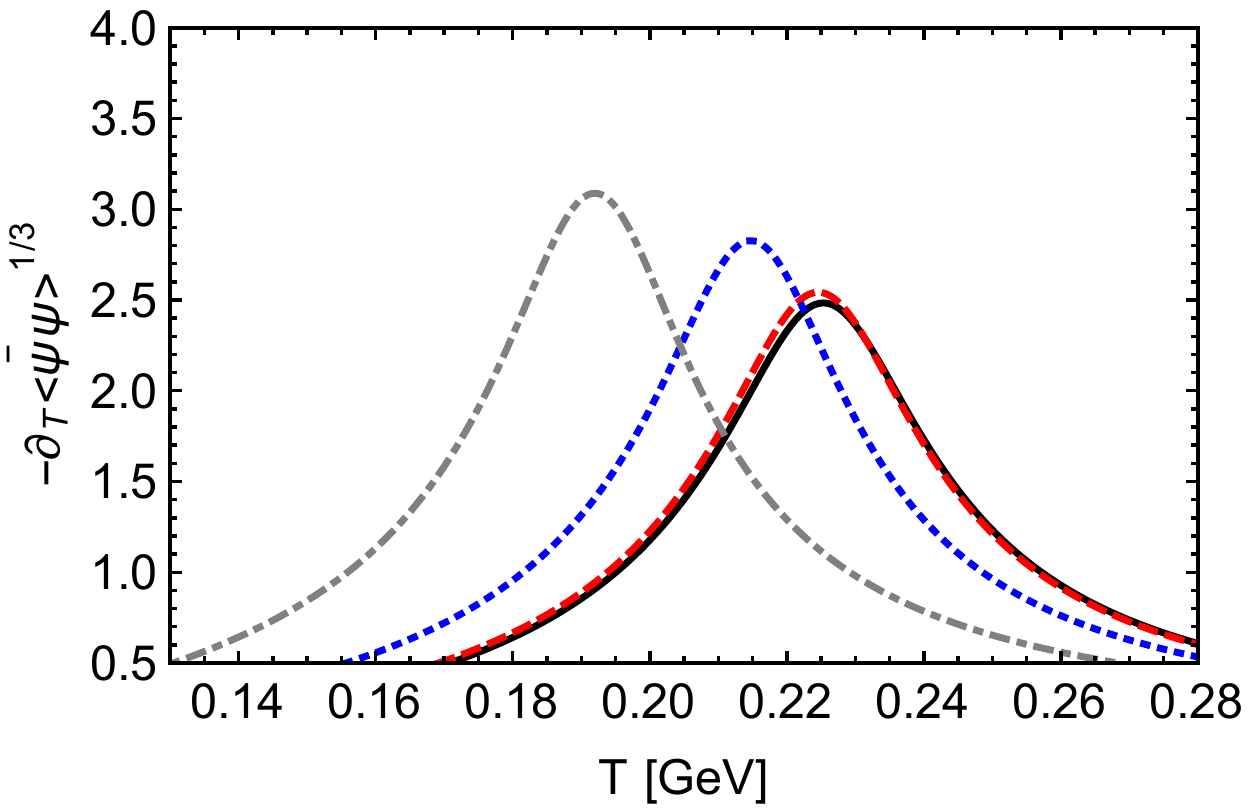}
\caption{Thermal grandient of the chiral condensate as a function of the temperature for different magnetic field strength. {\em Left panel:} $B-$independent coupling (Magnetic Catalysis), {\em Right panel}: $B-$dependent coupling (Inverse Magnetic Catalysis). Solid black curves is for $eB=0\ {\rm GeV}^2$, red dashed curve, $eB=0.2\ {\rm GeV}^2$, blue dotted curve, $eB=0.6\ {\rm GeV}^2$, and grey dot-dashed curve, $eB=1\ {\rm GeV}^2$. }
\label{Fig.6a}
\end{center}
\end{figure}

\begin{figure}[th!]
\begin{center}
\includegraphics[width=0.48\textwidth]{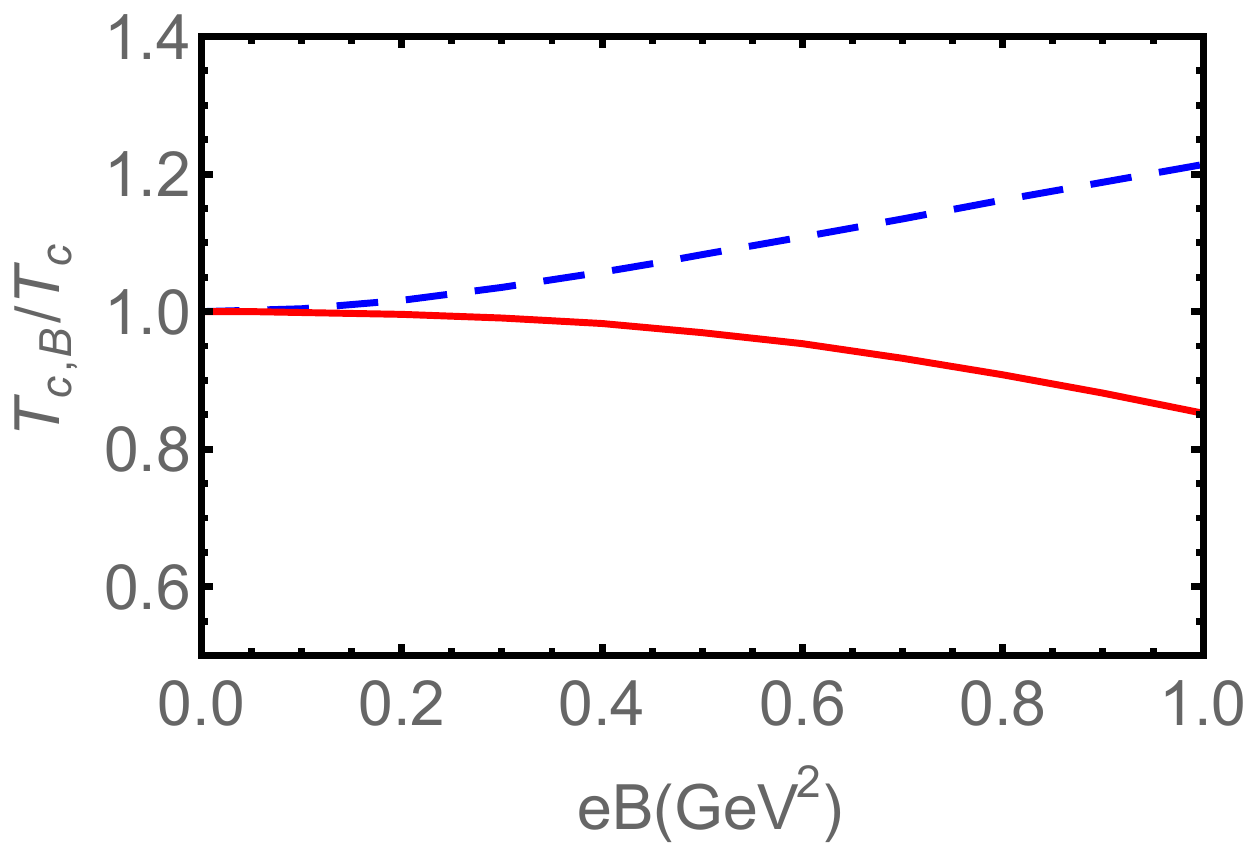}
\caption{Effective phase diagram in the $T-eB$ plane: Blue dashed curve corresponds to the constant $\alpha_{\rm eff}(0)$ in Eq.~(\ref{p1}), whereas the red solid curve is obtained with  $\alpha_{\rm eff}(\kappa)$ in Eq.~(\ref{coup}).}
\label{Fig.6}
\end{center}
\end{figure}

 The physical effect of the magnetic field must be considered in the coupling of the theory as well. This is so because on one hand, the gluon mass scale $m_G\sim \sqrt{\alpha_s |eB|}$, on one hand, while the strong coupling $\alpha_s\sim (b\ln(|eB|/\Lambda^2_{QCD}))^{-1}$ (with $b$ some constant number)~\cite{chinese,miransky,mueller,Ferreira}. For $|eB|$ of the order of the chiral transition temperatures, $\alpha_{\rm eff}\sim 1/|eB|$ and thus we propose the following functional form of the coupling which decreases with the magnetic field strength as
~\cite{Ferreira:2015gxa}
\begin{equation}
\alpha_{\rm eff}(\kappa)=\alpha_{\rm eff}(0) \bigg(\frac{1+a\kappa^{2}+b\kappa^{3}}{1+c\kappa^{2}+d\kappa^{4}}\bigg),\label{coup}
\end{equation}
with the parameters  
$\kappa=eB/\Lambda^2_{QCD}$, $a=0.002$, $b=-8.06\times 10^{-6}$, $c=0.004\times10^{-4}$, $ d=0.06\times10^{-4 }$ and we take $\Lambda_{QCD}=240$ MeV. In the spirit of Ref.~\cite{Ferreira:2015gxa}, the choice of parameters is made so as to reproduce the critical transition temperatures for the chiral and deconfinement transitions for different values of the magnetic field strength in the range 0-1GeV$^2$ obtained by lattice simulations, Ref.~\cite{Bali:2012zg}. The behavior of the coupling, normalized by the mean field value, Eq.~(\ref{p1}), is depicted in Fig.~\ref{Fig.4}.
The corresponding averaged 
condensate 
as a function of temperature and magnetic field strength is shown in Fig.~\ref{Fig.5}.  Again, we identify $T_{c,B}$  from the position of the peak of the thermal gradient of this quantity. Focusing on the behavior of the condensate, we observe that for $T<T_{c,B}$, the condensate rises with $eB$, whereas as $T\ge T_{c,B}$, it starts diminishing its magnitude with this scale, an unequivocal signal of inverse magnetic catalysis (see the second article in Ref.~\cite{Bali:2012zg}). The effect is better seen from the position of the peak of the thermal gradient of the condensate, as shown in Fig.~\ref{Fig.6a}. For a $B$-independent coupling, the corresponding $T_c$ increases as $eB$ grows bigger, as expected in the magnetic catalysis phenomenon. When the coupling diminishes with the magnetic field strength, the corresponding $T_c$ also diminishes, revealing the fingerprints of the Inverse Magnetic Catalysis. The corresponding  phase diagram is shown in Fig.~\ref{Fig.6}. Its behavior strongly resembles findings from lattice~\cite{Bali:2012zg} and other effective models approaches~\cite{Farias:2014, Ayala:2014gwa, Ferreira:2015gxa} with no ``turn over'' effect at intermediate $eB$~\cite{Ferreira}.


%

\section{Discussion and Conclusions}
We have studied the effective QCD phase diagram in the $T-eB$ plane within a confining contact interaction model. Such a model differs from the standard NJL theory by considering an infra-red cut-off which in addition to the dynamical ultra-violet scale, renders the quark propagator pole-less, hence supporting confinement. At finite temperature, we regularize the gap equation ensuring the coincidence of the chiral and confinement transitions at the same pseudo-critical temperature $T_{c}^\chi\simeq T_c^c\equiv T_c=225$~MeV. We then include the influence of a uniform magnetic field in the Schwinger proper-time formalism.

In the mean field limit, our effective phase diagram for the chiral transition shows an increasing of $T_{c,B}$ for strong magnetic fields. This picture is in agreement with the appearance of magnetic catalysis in our model. The rising behavior of $T_{c,eB}$ might be understood because a constant $\alpha_{\rm eff}(0)$ is fully oblivious to any reminiscent back reaction effect of gluon interacting with magnetic fields which later would have been integrated out to define in our model. On the contrary, the magnetic field dependent coupling  $\alpha_{\rm eff}(\kappa)$ of Eq.~(\ref{coup}) mimics the screening of gluon interactions in the infra-red that occurs in QCD and triggers the appearance of the inverse magnetic catalysis effect.  Physically, inverse magnetic catalysis is about the balance between the magnetic field and the temperature in the strength of interactions. At zero temperature, the gluon cloud that dresses the valence quarks is driving the phenomenon of magnetic catalysis, but at larger temperatures, gluons cease to have a prominent role and all that remains is the weakly coupled dynamics of quarks approaching their asymptotically free regime.
Our findings provide support to models in which the effective coupling, which may be considered proportional to the running coupling of QCD, behave as  monotonically decreasing functions of the strength of the external magnetic field, but extends over these models in the sense that the confinement-deconfinement transition is considered in terms of the evolution of the confining scale $\bar{\tau}_{ir}^{-1}$. By construction we have considered that the chiral and deconfinement transitions are coincidental at $B=0$, but have retained this hypothesis for finite magnetic field strength. Thus, our phase diagram actually describes the chiral symmetry breaking-restoration and confinement-deconfinement transitions. 
The running coupling model that we used in this work should be improved by including the effect of the temperature and eventually the baryon chemical potential for the more realistic description of the phase diagram. Rather than parameterizing the behavior of the effective coupling from lattice, an immediate goal is to determine the behavior of such a {\it running coupling} within the same framework. Furthermore, the entanglement between the chiral and confinement pseudo-critical temperatures already hints that the inverse magnetic catalysis also modifies the mechanism for confinement in a non trivial way, which is still worth to explore in further detail to complement our current understanding of the magnetic field influence on confinement without advocating the Polyakov loops.  Though a priori there is no first principles constraint that ensures or discards coincidence of the chiral and deconfinement transitions in the magnetized QCD vacuum, as we have assumed in this article, the evolution of the confining scale might well serve as a genuine parameter to explore the traits of the latter transition independently from the chiral order parameters. Also the role of anisotropies in the coupling and tensor interactions should be accounted for  to explore whether the order of the transition changes at larger magnetic fields. We are considering all  the above  and will report findings elsewhere. 

\ack
We acknowledge  A. Ayala, A. Bashir, G. Krein, A.~J. Mizher, Si-Xue Qin and C. Villavicencio  for valuable discussions. We also acknowledge the participants of the 5th International Workshop on Non Perturbative Aspect of Field Theory, Morelia 2015, for providing a valuable atmosphere for exchange of ideas, which lead to the genesis of this work. We further acknowledge support from CIC-UMSNH (Mexico) under project 4.22. AA acknowledges IFM-UMSNH, CONACyT (Mexico) and Department of Physics, Gomal University (Pakistan) for support. AR also acknowledges CONACyT (Mexico) support under grant 128534.

\end{document}